# Crystal chemistry and *ab initio* investigations of ultra-hard dense rhombohedral carbon and boron nitride

Samir F. Matar [1,§,*] and Vladimir L. Solozhenko [2]

[1] Lebanese German University (LGU), Sahel Alma, Jounieh, Lebanon
https://orcid.org/0000-0001-5419-358X

[2] LSPM–CNRS, Université Sorbonne Paris Nord, 93430 Villetaneuse, France
https://orcid.org/0000-0002-0881-9761

[§] *Former DR1-CNRS senior researcher at the University of Bordeaux, ICMCB-CNRS, France*

[*] Corresponding author email: s.matar@lgu.edu.lb and abouliess@gmail.com

**Abstract**

*Rhombohedral dense forms of carbon, rh-$C_2$ (or hexagonal h-$C_6$), and boron nitride, rh-BN (or hexagonal h-$B_3N_3$), are derived from rhombohedral 3R graphite based on original crystal chemistry scheme backed with full cell geometry optimization to minimal energy ground state computations within the quantum density functional theory. Considering throughout hexagonal settings featuring extended lattices, the calculation of the hexagonal set of elastic constants, provide results of large bulk moduli i.e. $B_0(rh-C_2)$ = 438 GPa close to that of diamond, and $B_0(rh-BN)$ = 369 GPa close to that of cubic BN. The hardness assessment in the framework of three contemporary models enables both phases to be considered as ultra-hard. From the electronic band structures calculated in the hexagonal Brillouin zones, 3R graphite is a small-gap semiconductor, oppositely to rh-$C_2$ that is characterized by a large band gap close to 5 eV, as well as the two BN phases.*

**Dedication:**

In memoriam of Prof. Gérard Demazeau (University of Bordeaux-France).  † November 2017

## 1- Introduction and crystal chemistry rationale

Graphite is mainly known as 2*H* structure with two carbon layers in hexagonal *P*6$_3$/*mmc* space group. Nevertheless, early crystal structure analyses of natural graphite by Lipson and Stokes [1, 2] showed the occurrence of extra lines in its X-ray pattern. The hypothesis of impurity was rapidly discarded, and the authors concluded to the presence of a non negligible percentage of a rhombohedral phase called 3*R* graphite in $R\bar{3}m$ space group. The structure shown in Fig. 1a using hexagonal axes (for a better presentation than with a rhombohedron) is characterized by 3 layers. Both hexagonal and rhombohedral graphite exhibit a triangular planar coordination of sp$^2$ carbon with an angle of 120°. In 2*H* graphite the carbon atoms occupying the two layers are in fixed Wyckoff positions, i.e. C1 at 2*b* (0, 0, ¼) and C2 at 2*c* (1/3, 2/3, ¼). Oppositely, in 3*R* graphite (three layers) there is one carbon equivalent at the special twofold 2*c* (*x, x, x*) Wyckoff position of $R\bar{3}m$ space group. By using hexagonal axes, the atoms are in the six-fold position 6*c* (0, 0, *z*). With *z* = 0.164 a closer agreement with the X-ray intensities was obtained, leading to a slight ±0.03Å out-of-plane displacement of atoms, i.e. with puckered layers whereas with *z* = 0.167 (1/6) a planar configuration is kept [1]. Then allowing changes in *z* with different magnitudes below 1/6 induces departures from the co-planarity of C atoms to increasingly puckered layers leading eventually to a 3D atomic arrangement alike in diamond as shown here below. A lowering of *z* from 0.164 down to 0.125 let obtain an intermediate structure shown in Fig. 1b accompanied by a change of ∠C-C-C angle from 120° characteristic of sp$^2$-carbon (Fig. 1a) down to 96°. The resulting constrained structure was then submitted to full geometry relaxation using quantum density functional theory (DFT) calculations [3, 4] (cf. next section), leading to ∠C-C-C = 109.47° characteristic of C(sp$^3$) and a smaller volume . The relaxed structure in Fig. 1c presents a succession of corner sharing *C4* tetrahedra in a diamond-like manner.

Extending the crystal chemistry mechanism to boron nitride, a similar approach was adopted with the difference of a lowering of the symmetry from $R\bar{3}m$ (space group No 166) *to R*3*m* (space group N$^o$ 160) due to the occupation of the 2*c* (*x, x, x*) position - or 6*c* (0, 0, *z*) -, by two different chemical species, B and N. A rhombohedral BN predicted from *ab initio* simulated annealing was cited in 2008 [5]. The structure shown in Fig. 2i was subsequently calculated with ground state energy close to the new 3*R* graphitic BN (Table 1) obtained from the carbon analogue (Fig. 1a). However we did not retain it as model for present study due to the large difference of structural arrangement of the layers with respect to the model derived from 3*R* C$_2$ which leads to 3*R* BN sketched in Fig. 2a. This model followed in the representations in Fig. 2b, and Fig. 2c indicates a mechanism of structure transformation closely resembling the one used for carbon.

## 2- Computational framework

The search for C and BN ground state structures and calculations of energies of the different phases were based on DFT within the plane-wave Vienna Ab initio Simulation Package VASP code [6, 7] used with the projector augmented wave (PAW) method [7,8] for the atomic potentials with all valence states, especially, in regard of such light elements as B and N. The effects of exchange-correlation were accounted for calling for the generalized gradient approximation (GGA) [9]. A conjugate-gradient algorithm according to Press *et al*. [10] was used in this computational scheme to relax the atoms onto the ground state. The tetrahedron method with Blöchl *et al*. corrections [11] and Methfessel-Paxton scheme [12] were applied for both geometry relaxation and total energy calculations. Brillouin-zone (BZ) integrals were approximated using a special **k**-point sampling of Monkhorst and Pack [13]. The optimization of the structural parameters was performed until the forces on the atoms were less than 0.02 eV/Å, and all stress components less than 0.003 eV/Å$^3$. The calculations were converged at an energy cut-off of 500 eV for the plane-wave basis set concerning the **k**-point integration in the Brillouin zone, with a starting mesh of 6×6×6 up to 12×12×12 for best convergence and relaxation to zero strains. In the post-treatment process of the ground state electronic structures, the charge density and the electronic band structures are computed and illustrated. The mechanical properties were derived from the calculation of the elastic constants $C_{ij}$ in the hexagonal system.

## 3- Calculations and results.

3-1. *Crystal structures using hexagonal setting.*

From full geometry relaxation and ground state energies, Table 1 provides the resulting crystal parameters and the corresponding energies for all $C_6$ and $B_3N_3$ structures. For graphitic $C_6$ the experimental [1] and calculated data provided in the first two columns exhibit slight differences for *a* and *c* lattice constants, especially, for a larger calculated *c*. The starting $z_C = 0.164$ changes to 1/6, thus showing that planar carbon arrangement is favored energetically *versus* puckered layers structure. Agreement with experiment is also found for the interatomic distance d(C-C) = 1.42 Å.

Confronted with 2*H* graphite, the total energy for 4 C (note *i* at the bottom of Table 1) is very close to 3*R* C2 total energy. 3D *rh*-$C_2$ on the other hand is characterized by a slightly higher total energy, and its energy is equal to that of diamond and is less than that of lonsdaleite (hexagonal diamond) – indicated in note *j*.

Turning to BN phases, the same trend of slightly higher 3D *rh*-BN *versus* 3*R* BN is noted, the latter being at energy close to 3*R* BN [5] – note *k*. The total energy values are mirrored by the trends of cohesive energies given at the last line of Table 1. Regarding the crystal structures, the 2D phases are characterized by systematically larger $c$ hexagonal parameters whilst $a$ planar lattice constant changes much less, and the 2D phases are characterized by $z_{C;B;N}$ = 0.167 / 0.833, while the 3D phases have $z_{C;B;N}$ = 0.125 / 0.875. Finally, the shortest distances are smaller in 2D phases due to in-plane binding within the layer, and larger in 3D phases where the distance describes the *C4* tetrahedron (Fig. 1c) or *BN3* / *NB3* tetrahedra in Fig. 2c.

3-2. *Charge density*

To further assess the electronic and crystal structure relationship, the charge density projections onto the chemical constituents as situated in the crystal lattice are needed. Fig. 3a shows the charge density (yellow) volumes in 3*R* graphite expressed in three layers within continuous yellow rings of sp$^2$-like C. Oppositely in Fig. 3b the charge density around carbon atoms in *rh*-C$_2$ (C$_6$ in hexagonal coordinates) shows a perfect sp$^3$ tetrahedral shape carbon charge density arranged into tetrahedra reproducing the structure sketch in Fig. 1c.

Clearly, the two carbon compounds are perfectly covalent chemical systems because of the presence of one chemical species, C, with electronegativity χC = 2.55. The situation becomes less obvious upon accounting for BN where χB = 2.04 is much lower than χN = 3.04. The larger electronegativity of N leads to a charge transfer B→N resulting in charge-poor B and charge-rich N as illustrated in next paragraph. The 3*R* BN charge density in Fig. 3c illustrates the polar covalent chemical behavior with the charge pointing from N towards B with an ovoid shape especially visible in the middle layer. In Fig. 3d showing the charge density of 3D *rh*-BN, the tetrahedral arrangement which is similar to Fig. 3b of 3D carbon analogue is visible on N charge density, especially, in the *B3N* tetrahedron at the lower part of the figure.

3-3. *Electronic band structures*

Figure 4 shows the electronic band structures along the main directions of the first wedge in rhombohedral Brillouin zones. All four carbon phases possess band gaps separating the filled valence band (VB) from the empty conduction band (CB), so that the zero of energy along the *y*-axis is at $E_V$ top of the valence band.

In 2D carbon, the gap is very small in agreement with a semi-conducting character. Oppositely, the 3D carbon phase has a large band gap, indirect, with a value close to 5 eV. This is also observed in the 3D BN phase with a slightly lower band gap of 5 eV. The iono-covalent character brought by the presence of two chemical elements with different electronegativities is exhibited in the band structure of 3D BN where the s-like bands due to nitrogen are below -15 eV and well separated from the p block above this energy. Both 3D phases are considered as large band gap insulators, one covalent and the other iono-covalent.

3-4. *Mechanical properties.*

The assessment of the mechanical properties is based on the elastic properties determined by performing finite distortions of the lattice and deriving the elastic constants from the strain-stress relationship.

In hexagonal symmetry there are six independent elastic stiffness constants $C_{11}$, $C_{33}$, $C_{44}$, $C_{12}$, and $C_{13}$. Most compounds are encountered in polycrystalline form where one may consider the single crystalline grains randomly oriented. Consequently on a large scale, such materials can be considered as statistically isotropic. They are then fully described by the bulk modulus $B$ and the shear modulus $G$ that may be obtained by averaging the single-crystal elastic constants. A widely used averaging method of elastic constants is Voigt's one [14] based on a uniform strain. The calculated set of elastic constants (in GPa) are:

*rh*-C$_2$      $C_{11} = 1117$;    $C_{12} = 93$;    $C_{13} = 61$;    $C_{33} = 1280$;    $C_{44} = 510$.

*rh*-BN      $C_{11} = 915$;    $C_{12} = 119$;    $C_{13} = 74$;    $C_{33} = 960$;    $C_{44} = 352$.

All $C_{ij}$ values are positive and their combinations: $C_{11} > C_{12}$, $C_{11}C_{33} > C_{13}^2$ and $(C_{11}+C_{12})C_{33} > 2C_{13}^2$ obey the rules pertaining to the mechanical stability of the phase.

The bulk ($B_V$) and shear ($G_V$) moduli following Voigt are formulated as:

$$B_V = 1/9 \{2(C_{11} + C_{12}) + 4C_{13} + C_{33}\},$$

and

$$G_V = 1/30 \{C_{11} + C_{12} + 2\,C_{33} - 4C_{13} + 12\,C_{44} + 6(C_{11} - C_{12})\}$$

The calculated moduli are then as follows:

*rh*-C$_2$:    $B_V = 438$ GPa,  $G_V = 542$ GPa

*rh*-BN:    $B_V = 369$ GPa,  $G_V = 389$ GPa

These results are in close agreement with corresponding values of diamond and *c*-BN (see Tables 2b and 3b).

Vickers hardness ($H_V$) was predicted using three contemporary theoretical models of hardness: (i) Mazhnik-Oganov model [15], (ii) Chen-Niu model [16], and (iii) thermodynamic model [17]. The first two models use the elastic properties, while the thermodynamic model is based on crystal structure and thermodynamic properties. Mazhnik-Oganov model was also used for the estimation of fracture toughness ($K_{Ic}$). The results are presented in Tables 2 and 3.

Tables 2a and 3a present Vickers hardness ($H_V$) and bulk moduli ($B_0$) calculated in the framework of the thermodynamic model of hardness, and Tables 2b and 3b present the other mechanical properties such as shear modulus ($G$), Young's modulus ($E$), the Poisson's ratio ($v$) and fracture toughness ($K_{Ic}$).

A slightly higher hardness of rhombohedral carbon compared to diamond (both cubic and hexagonal) is observed for all three models (Table 2a). The similar trend is also found for the dense rhombohedral boron nitride *versus* cubic and wurtzite polymorphs. A good agreement is observed for the bulk moduli estimated using the thermodynamic model ($B_0$) with the values calculated from the set of elastic constants ($B_V$).

In general, both *rh*-$C_2$ and *rh*-BN have superior mechanical properties and thus can be considered as prospective ultra-hard materials [28].

**4- Conclusion**

Crystal chemistry rationale of structurally transforming rhombohedral graphitic carbon and boron nitride into three dimensional dense forms was followed by geometry-optimization calculations within DFT. The ground state energies of 3D structures of *rh*-$C_2$ (or *h*-$C_6$) and *rh*-BN (or *h*-$B_3N_3$) were determined as well as the energy-related physical quantities as elastic constants, charge density and electronic band structures. Compared to diamond and lonsdaleite, *rh*-$C_2$ is shown to possess slightly better mechanical properties such as hardness and fracture toughness. Also, *rh*-BN shows hardness value close to those of superhard cubic boron nitride.

Table 1  Hexagonal setting of rhombohedral carbon (SG #166) and boron nitride (SG #160).
All atoms are at (0, 0, $z$) positions. Distances are in Å and energies in eV units.

|  | $C_6$ | | | $B_3N_3$ | |
|---|---|---|---|---|---|
|  | 2D (exp) | 2D | 3D | 2D | 3D |
| $a$ | 2.52 | 2.46 | 2.52 | 2.51 | 2.56 |
| $c$ | 10.21 | 10.64 | 6.17 | 9.65 | 6.27 |
| $z_C$ | 0.164 | 0.167 | 0.125 | | |
| $z_B$ | | | | 0.167 | 0.125 |
| $z_N$ | | | | 0.833 | 0.875 |
| $d_{C-C\,(B-N)}$ | 1.42 | 1.42 | 1.54 | 1.45 | 1.57 |
| $E_{tot}$ | | -55.32 | -54.54 | -52.59 | -52.26 |
| $E_{tot}$/at | | -9.22 [i] | -9.09 [j] | -8.76 [k] | -8.74 [l] |
| $E_{coh}$/at | | -2.75 | -2.62 | -2.58 | -2.52 |

Comparative energies (eV/at.):

[i] graphite 2*H*: -9.32
[j] diamond: -9.09, lonsdaleite: -9.06
[k] $B_3N_3$ [5]: -8.78
[l] c–BN: -8.71

Atomic energies: $E$ (C) = -6.47 eV , $E$ (B) = -5.57 eV, $E$ (N) = -6.8 eV

Table 2a  Vickers hardness ($H_V$) and bulk moduli ($B_0$) of carbon allotropes calculated in the framework of the thermodynamic model of hardness [17]

|  | Space group | $a$ (Å) | $c$ (Å) | $\rho$ (g/cm$^3$) | $H_V$ (GPa) | $B_0$ (GPa) |
|---|---|---|---|---|---|---|
| $rh$-C$_2$ | $R\text{-}3m$ | 2.5190 | 6.17026 | 3.5294 | 98 | 445 |
| $rh$-C$_4$ [18] | $R\text{-}3m$ | 2.5004 | 12.2100 | 3.6204 | 100 | 456 |
| Lonsdaleite | $P6_3/mmc$ | 2.5221[†] | 4.1186[†] | 3.5164 | 97 | 443 |
| Diamond | $Fd\text{-}3m$ | 3.56661[‡] | – | 3.5169 | 98 | 445[§] |

[†] Ref. 19
[‡] Ref. 20
[§] Ref. 21

Table 2b  Mechanical properties of carbon allotropes: Vickers hardness ($H_V$), bulk modulus ($B_0$), shear modulus ($G$), Young's modulus ($E$), Poisson's ratio ($\nu$) and fracture toughness ($K_{Ic}$)

|  | $H_V$ | | | $B$ | | $G$ | $E$[§] | $\nu$[§] | $K_{Ic}$[†] |
|---|---|---|---|---|---|---|---|---|---|
|  | T[*] | MO[†] | CN[‡] | $B_0$[*] | $B_V$ | | | | |
|  | GPa | | | | | | | | MPa·m$^{½}$ |
| $rh$-C$_2$ | 98 | 103 | 99 | 445 | 438 | 542$^{(G_v)}$ | 1151 | 0.062 | 6.4 |
| $rh$-C$_4$ [18] | 100 | 105 | 97 | 456 | 458 | 552$^{(G_v)}$ | 1181 | 0.070 | 6.7 |
| Lonsdaleite | 97 | 99 | 94 | 443 | 432 | 521$^{(G_v)}$ | 1115 | 0.070 | 6.2 |
| Diamond | 98 | 100 | 93 | 445[**] | | 530[**] | 1138 | 0.074 | 6.4 |

[*] Thermodynamic model [17]
[†] Mazhnik-Oganov model [15]
[‡] Chen-Niu model [16]
[§] $E$ and $\nu$ values calculated using isotropic approximation
[**] Ref. 21

Table 3a  Vickers hardness ($H_V$) and bulk moduli ($B_0$) of dense BN polymorphs calculated in the framework of the thermodynamic model of hardness [17]

| | Space group | $a$ (Å) | $c$ (Å) | $\rho$ (g/cm$^3$) | $H_V$ (GPa) | $B_0$ (GPa) |
|---|---|---|---|---|---|---|
| *rh*-BN | *R-3m* | 2.54365 | 6.24479 | 3.5361 | 56 | 382 |
| *rh*-B$_2$N$_2$ [22] | *R-3m* | 2.5004 | 12.24821 | 3.4691 | 55 | 375 |
| *w*-BN | *P6$_3$/mc* | 2.5505[†] | 4.210[†] | 3.4756 | 54 | 374 |
| *c*-BN | *F-43m* | 3.6160[‡] | – | 3.4869 | 55 | 377[§] |

[†] Ref. 23
[‡] Ref. 24
[§] Ref. 25

Table 3b  Mechanical properties of dense BN polymorphs: Vickers hardness ($H_V$), bulk modulus ($B_0$), shear modulus ($G$), Young's modulus ($E$), Poisson's ratio ($\nu$) and fracture toughness ($K_{Ic}$)

|  | $H_V$ | | | $B$ | | $G$ | $E$ [§] | $\nu$ [§] | $K_{Ic}$ [†] |
|---|---|---|---|---|---|---|---|---|---|
|  | T[*] | MO[†] | CN[‡] | $B_0$[*] | $B_V$ | | | | |
|  | GPa | | | | | | | | MPa·m^½ |
| *rh*-BN | 56 | 72 | 67 | 382 | 369 | 389 [(Gv)] | 864 | 0.110 | 4.6 |
| *rh*-B$_2$N$_2$ [22] | 55 | 75 | 73 | 375 | 356 | 400 [(Gv)] | 873 | 0.091 | 4.4 |
| *w*-BN | 54 | 70 | 64 | 374[**] | | 384[††] | 858 | 0.118 | 4.7 |
| *c*-BN | 55 | 74 | 69 | 377[**] | | 402[‡‡] | 890 | 0.107 | 4.8 |

[*] Thermodynamic model [17]

[†] Mazhnik-Oganov model [15]

[‡] Chen-Niu model [16]

[§] $E$ and $\nu$ values calculated using isotropic approximation

[**] Ref. 25

[††] Ref. 26

[‡‡] Ref. 27

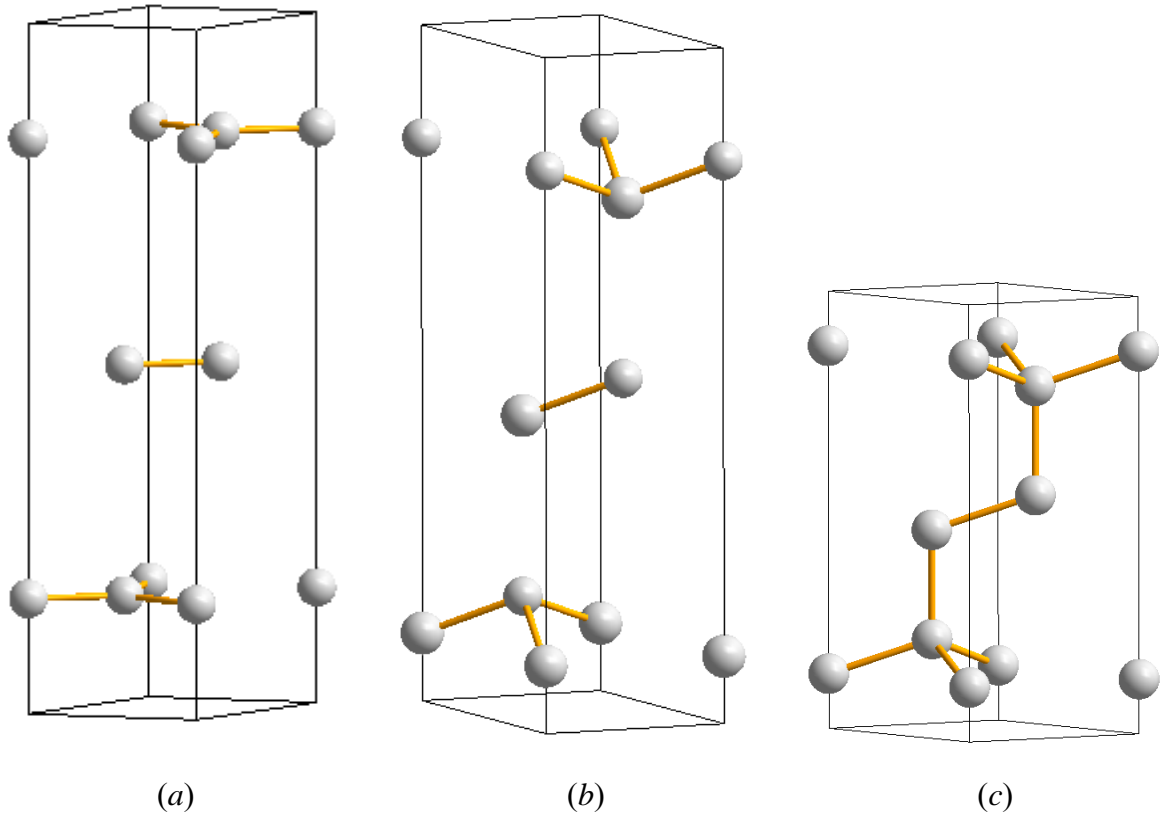

Figure 1. Transformation mechanism using hexagonal setting of rhombohedral structures. (*a*) 2D 3*R* graphite with C(sp$^2$), $z$(C) ~ 0.167; (*b*) with $z$(C) = 0.125 a non-relaxed intermediate phase is obtained with three non-planar puckered layers; (*c*) fully relaxed 3D new phase with tetrahedral sp$^3$ carbon alike diamond resulting into a much more compact structure with smaller *c* hexagonal parameter. The ∠C–C–C angle changes from (*a*) 120° (sp$^2$) to (*b*) 96°, and (*c*) 109.47° (sp$^3$).

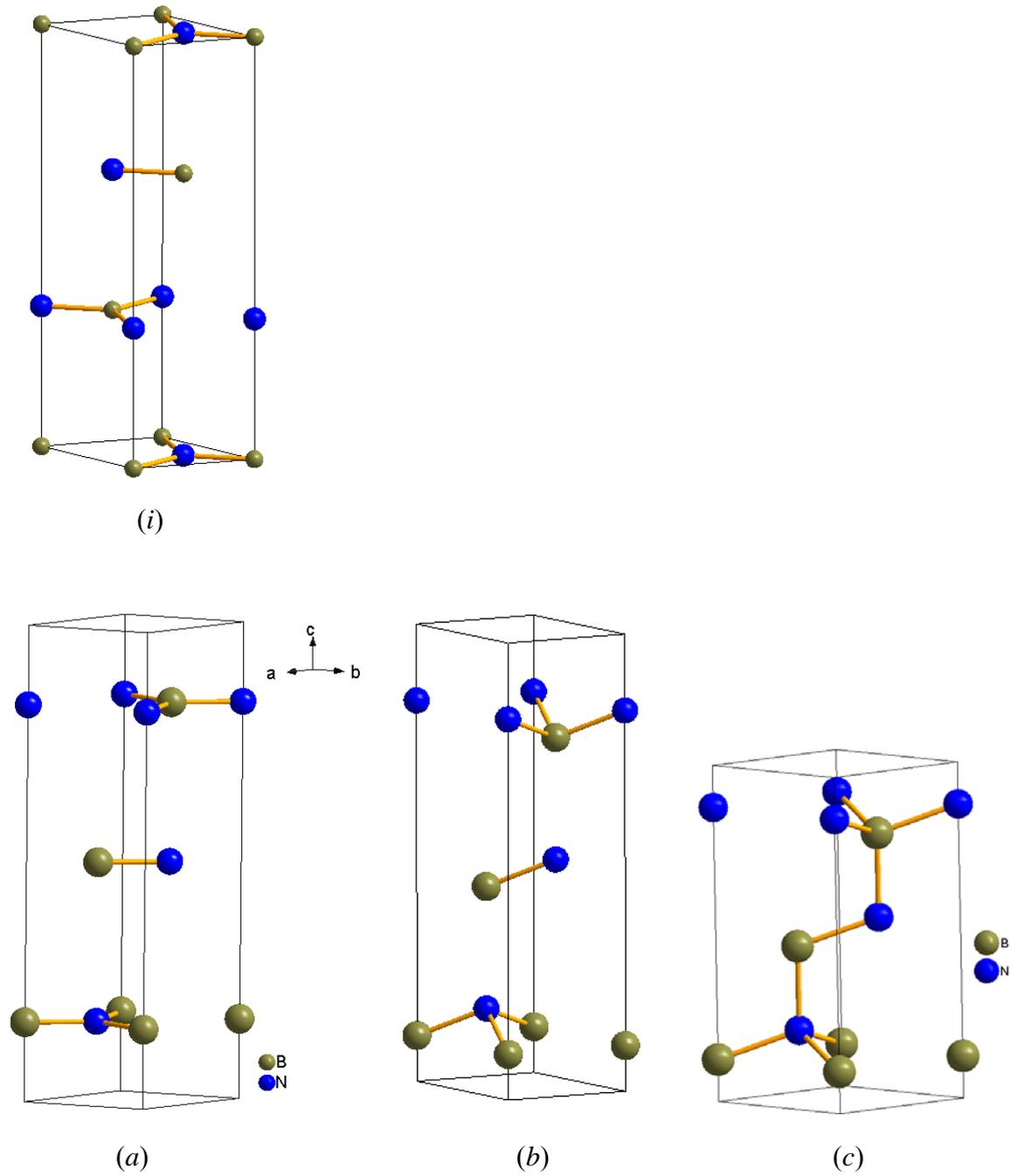

Figure 2. Transformation mechanism using hexagonal setting of rhombohedral BN structures in hexagonal settings. (*i*) *rh*-BN 3R from Ref. [5], not retained herein as model. (*a*) $B_3N_3$ with $z(B) = 0.167$ and $z(N) = 0.833$; (*b*) non-relaxed intermediate phase with three non-planar layers, and (*c*) fully relaxed 3D new BN ($B_3N_3$) with *BN3* and *NB3* tetrahedra resulting into a much more 3D compact structure.

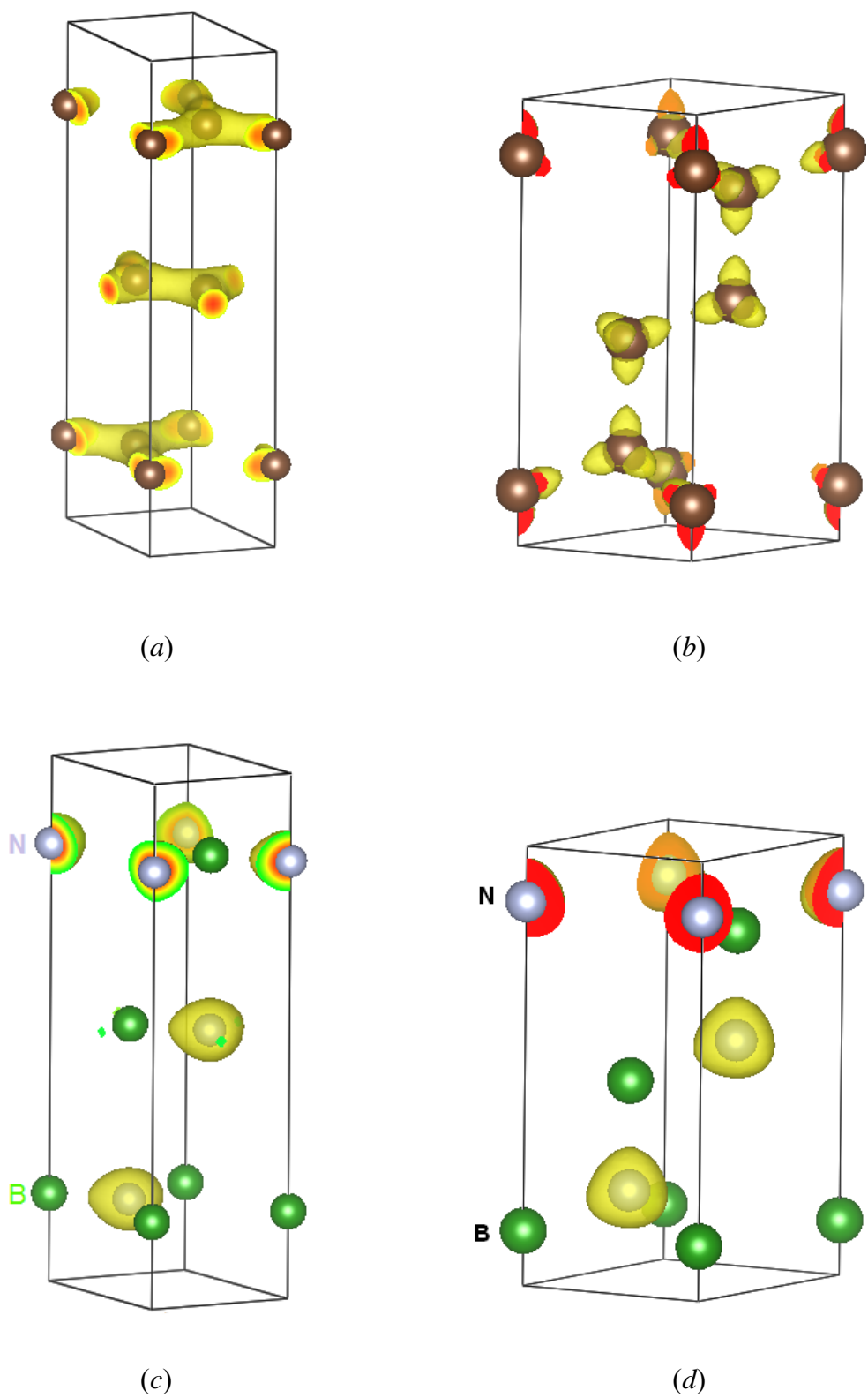

Figure 3. Charge densities of the rhombohedral structures in hexagonal settings. (*a*) $C_6$ 3*R*, (*b*) 3D $C_6$, (*c*) $B_3N_3$ 3*R*, (*d*) 3D $B_3N_3$

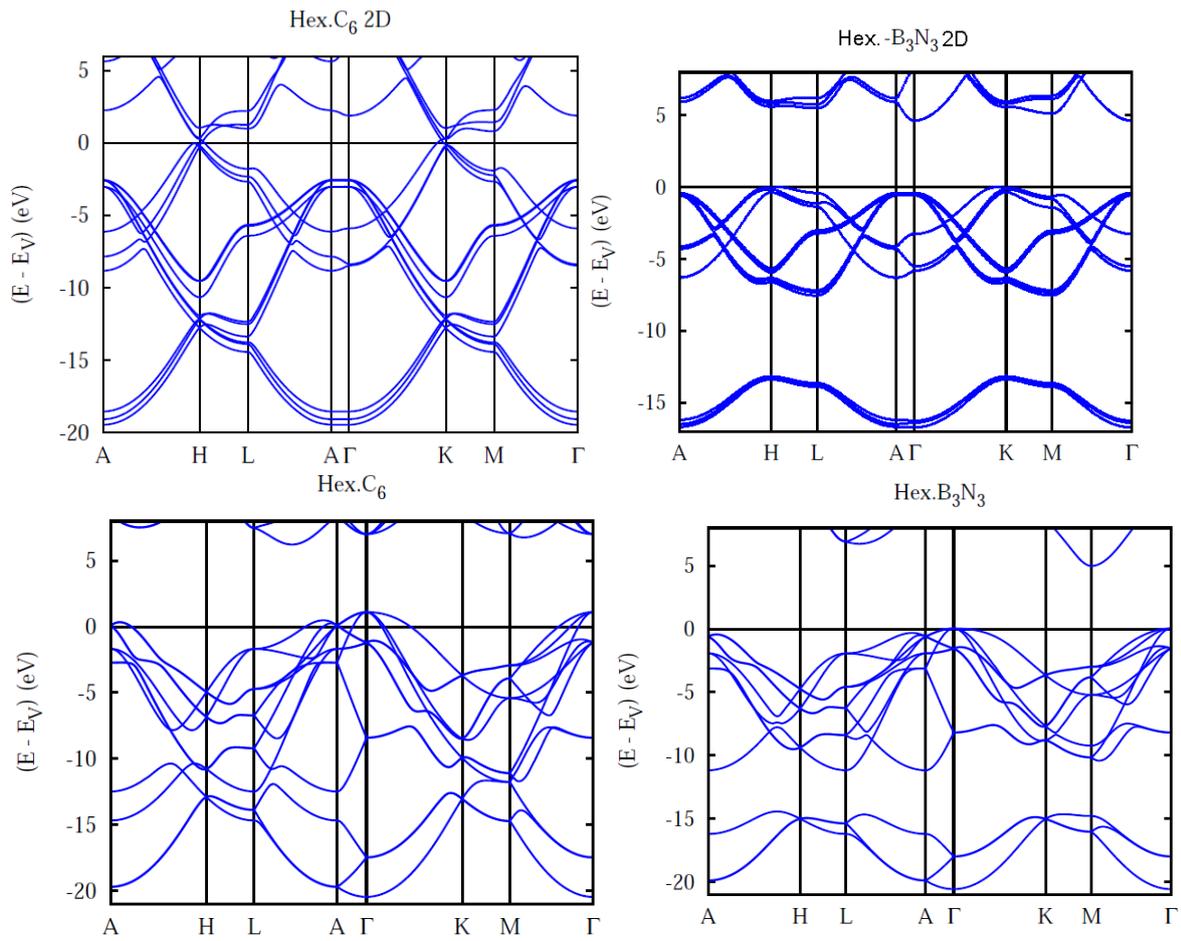

Figure 4.  Electronic band structures of 2D (top) and 3D (bottom) new rhombohedral forms of carbon and boron nitride.